# Deviations from mean-field behavior in disordered nanoscale superconductor–normal-metal–superconductor arrays

Taejoon Kouh and J. M. Valles, Jr.*
*Department of Physics, Brown University, Providence, Rhode Island 02912*

We have fabricated quasi-two-dimensional disordered arrays of nanoscale Pb grains coupled by an overlayer of Ag grains. Their temperature-dependent resistive transitions follow predictions for an array of mesoscopic superconductor–normal-metal–superconductor junctions. The decrease of their transition temperatures with Ag overlayer thickness systematically deviates from the Cooper limit theory of the proximity effect as the Pb grain size decreases. The deviations occur when the estimated number of Cooper pairs per grain is <1 and suggest the approach to a superconductor-to-metal transition.

Superconductor (S) normal-metal (N) structures are fundamentally interesting as systems with spatially varying order parameters whose properties can be successfully described using mean-field microscopic theories.[1–3] For example, "proximity effect" models account for the upper critical field of SN multilayers,[4] the spatial dependence of the tunneling density of states in mesoscopic SN structures,[5] and the critical currents of SNS junctions.[6] These proximity effects are simplest to describe in the so-called Cooper limit in which the dimensions of the S and N regions are smaller than $\xi_S$ and $\xi_N$, the S and N coherence lengths, respectively.[1,3,7] In this regime, the Cooper pair density is spatially uniform and the mean-field transition temperature $T_{co}$ is determined by the volume average of the pairing interaction strength over the N and S regions.[1,3] This approximation improves as the N and S dimensions decrease. Studies of nanoscale SN devices[8] and isolated ultrasmall superconducting grains,[9] however, suggest reasons that a mean-field description of SN composites might eventually break down in the extreme Cooper limit. They demonstrate that mesoscopic fluctuation and energy level quantization effects falling outside the realm of mean-field theory increasingly dominate their behavior as the size of the S and N regions decrease.

Recently, dramatic, non-mean-field behavior has been predicted for two-dimensional (2D) arrays of ultrasmall superconducting grains embedded in a metal.[10,11] Spivak, Zyuzin, and Hruska[11] presented evidence that these arrays undergo a superconductor-to-metal quantum phase transition (SMT) as the superconducting grain radius $r$ or concentration $x_{sc}$ decreases. This prediction deviates strongly from the Cooper limit model which would predict that $T_{co}$ only exponentially approaches zero as $r$ and $x_{sc}$ decrease. We present an investigation of this interesting prediction using disordered 2D arrays of nanoscale Pb grains coupled through an overlayer of Ag. By changing the Ag overlayer thickness $d_{Ag}$ at fixed Pb layer thickness $d_{Pb}$, we have been able to tune $x_{sc}$. By changing $d_{Pb}$ we have been able to change $r$. We find that $T_{co}$ of bilayers with the largest $r$ decreases exponentially with $x_{sc}$ in quantitative agreement with theory. Systematic deviations from theory appear and grow as $r$ decreases. They appear where fluctuations in the order-parameter amplitude on the grains are expected to be large and near the estimated critical concentration for the SMT.[11]

The nanoscale SNS arrays were fabricated using the structure that spontaneously forms in films that are quench condensed from vapor onto cryogenically cooled substrates. *In situ* scanning tunneling microscopy (STM) experiments have shown that quench condensed Pb films with bulk equivalent thicknesses $d_{Pb}$ up to $\simeq 3$ nm form a 2D disordered array of physically separate nanoscale grains.[12,13] At $d_{Pb}=3$ nm, the average grain radius, height, and intergrain gaps are 10, 4, and 1.2 nm, respectively. In thicker films, an overlayer of grains forms that bridge the intergrain gaps. The SNS "arrays" were made by first depositing Pb (superconductor) followed by a series of Ag (normal) depositions onto fire polished glass substrates held at 8 K. *In situ* STM on other Pb/Ag structures at 77 K reveal that the Ag forms an overlayer of grains that "bridge" the underlying Pb grains.[14] Each series of depositions and measurements was done in the ultrahigh vacuum environment of a dilution refrigerator without breaking vacuum. The equivalent bulk density film thicknesses were determined to an accuracy of 0.01 nm using a quartz-crystal microbalance. Predeposited Au/Ge pads provided electrical contact to the films and film sheet resistances $R$ were measured in their Ohmic regime using standard four-probe dc or ac techniques. Film homogeneity was checked by comparing the $R$ of adjacent film regions, which typically agreed to better than 5%. Data from three arrays with $d_{Pb}=1.5$, 2.2, and 3.0 nm are presented here. They are consistent with less complete data sets on separate arrays with both smaller and larger $d_{Pb}$.

The normal-state sheet resistance $R_N=R(8K)$ of the bilayers evolves with total film thickness in a manner resembling pure granular film systems [Fig. 1(a)].[12,15] The bilayer with $d_{Pb}=1.5$ nm became electrically continuous at $d_{Pb}+d_{Ag}=4.7$ nm, which falls between the thicknesses at which pure Ag (2.2 nm) and pure Pb films (5.5 nm) become continuous. All sets of films exhibited proximity effect driven insulator to superconductor transitions [e.g., Fig. 1(b)].[16,17] For the highest $R_N$ in Fig. 1(b) $R(T)$ shows a dip near 1.3 K, which is a signature of the appearance of local superconductivity and a consequence of the granular morphology of these films.[18] The initial superconducting transition is quite broad.

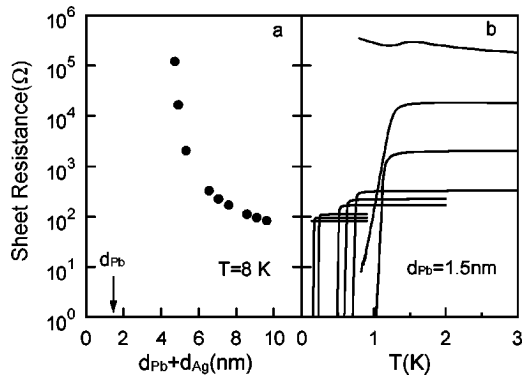

FIG. 1. (a) Sheet resistance at $T=8$ K vs total film thickness, $d_{Pb}+d_{Ag}$, with $d_{Pb}=1.5$ nm. The arrow indicates the thickness of the initial Pb film. (b) Sheet resistance vs temperature for the films in (a).

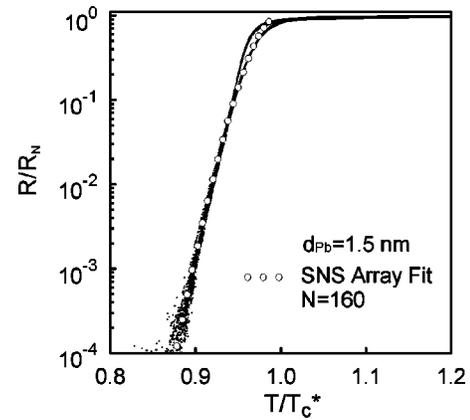

FIG. 2. Normalized sheet resistance vs reduced temperature $T/T_c^*$ of five of the Pb/Ag films shown in Fig. 1 ($d_{Pb}=1.5$ nm and $d_{Ag}=5, 5.5, 6.1, 7.1, 7.6$ nm). The open circles are a fit to the data using an array of mesoscopic SNS junctions model with $N$ transverse modes per junction.

Subsequent transitions become sharper and take place at lower $T$. Each occurs in a single step indicating that the Pb and Ag constituents simultaneously become superconducting.

Strong evidence that these bilayers form nanoscale SNS arrays comes from comparing their $R(T)$ to predictions for an array of mesoscopic SNS junctions. In general, $R(T)/R_N=[I_0(C\gamma/2)]^{-2}$ for a junction array, where $I_0$ is the modified Bessel function, $\gamma=hI_c(T)/ek_BT$ is the normalized energy barrier to a phase slip across a single junction, $I_c(T)$ is the junction critical current, $C\simeq 0.1$, and $R_N$ is the normal-state resistance.[19,20] For junctions with dimensions less than $\xi_N=(hD/k_BT)^{-1/2}$, where $D$ is the electronic diffusivity in the metal, $I_c(T)=Ne\Delta/h$, where $N=A/\lambda_F^2$ is the number of transverse modes in the point contact and $\Delta$ is the energy gap.[21] $A$ is the point contact area and $\lambda_F$ is the Fermi wavelength. For the bilayers, $\xi_N\simeq 100$ nm at 1 K based on the $R(d_{Ag})$ data, which exceeds the characteristic dimensions of the films. Thus the prediction for $R(T)/R_N$ depends only on $T_{co}$ through $\Delta$ and $N$.

The SNS array model fits the $R(T)$ of bilayers far from the insulator to superconductor transition over more than three decades. This agreement is shown in Fig. 2 for five of the $R(T)$ from Fig. 1(b). The data collapse onto a single curve when plotted as $R(T/T_c^*)/R_N$ where $R_N$ is the normal-state resistance and $T_c^*$ is within 3% of the midpoint temperature of each of the transitions. This process sets the parameter $T_{co}$. The resulting trajectory is well fit with $N=160\pm10$ and presuming that $\Delta$ has the BCS temperature dependence.[6] It is reasonable that $N$ assumes the same value for all five films. It implies that the Pb-Ag grain contact area does not change above the $d_{Ag}$ at which electrical continuity is established and agrees with our understanding of the film growth. $N=160$ corresponds to an average point contact area of $A=(4.8$ nm$)^2$. Similar fits yielded $N\simeq 400$, $A\simeq (7.5$ nm$)^2$ and $N\simeq 800$, $A\simeq (10.7$ nm$)^2$ for the $d_{Pb}=2.3$ nm and $d_{Pb}=3.0$ nm arrays, respectively. These areas are comparable to, but less than, the dimensions of the Pb grains for $d_{Pb}=3$ nm.[12,13] The increase in $A$ with $d_{Pb}$ suggests that the Pb grain size increases with $d_{Pb}$.

The reduction of the resistive midpoint temperature $T_{co}$ with increasing $d_{Ag}$ is shown in Fig. 3 for three bilayer sets. The uncertainties are smaller than the size of the symbols. For $d_{Pb}=3.0$ nm, $T_{co}(d_{Ag})$ is exponential over more than a decade. For $d_{Pb}=2.3$ nm and $d_{Pb}=1.5$ nm, $T_{co}$ exhibits a similar exponential decrease at high $T_{co}$, but appears to decrease more rapidly at lower $T_{co}$.

An exponential form of $T_{co}(d_{Ag})$ is expected for SN composites with low resistance SN interfaces and characteristic dimensions that are less than $\xi_S$ and $\xi_N$.[1,3,7,22] In this so-called Cooper limit, the effective superconducting coupling constant, $\lambda$, in the expression $T_{co}=T_o\exp(-1/\lambda)$ becomes the average of the coupling constants in the S and N regions. For the Pb-Ag bilayers $\lambda=\lambda_{Pb}d_{Pb}+\beta\lambda_{Ag}d_{Ag}/(d_{Pb}+\beta d_{Ag})$ where $\lambda_{Pb}$ and $\lambda_{Ag}$ are the coupling constants in the Pb and Ag regions, respectively, and $\beta$ is the ratio of the densities of states at the Fermi level of the Ag and Pb, $\beta = \nu_{Ag}/\nu_{Pb}$. If $\lambda_{Ag}=0$ then,

$$T_{co} = T_o \exp\left(-\frac{d_{Pb}+\beta d_{Ag}}{\lambda_{Pb} d_{Pb}}\right) \quad (1)$$

and $T_{co}$ decreases exponentially with $d_{Ag}$.[1,3,7] For a nonzero $\lambda_N$, small deviations from a simple exponential form should be apparent.[1]

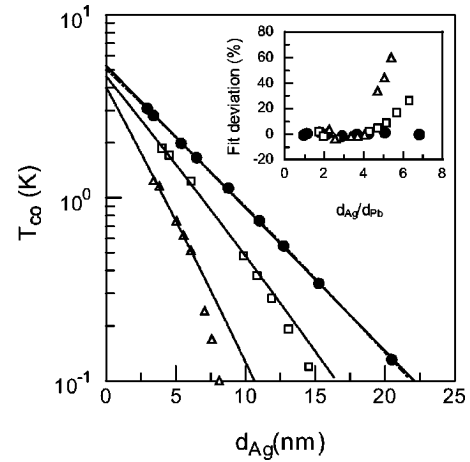

FIG. 3. Mean-field transition temperature of three Pb/Ag bilayer sets [$d_{Pb}=3.0$-nm (circles), 2.3 nm (squares), 1.5 nm (triangles)] as a function of Ag thickness. The $d_{Pb}=3$-nm data are fit to proximity effect theory in the Cooper limit with $\lambda_N=\lambda_{Ag}=0$ (solid line) and $-0.017$ (dashed line). The other solid lines are fits to the higher $T_{co}$ points of the $d_{Pb}=2.2$ and 1.5 nm bilayers by adjusting $T_o$ (see text). Inset: Fractional deviation of the data from the fits, $(T_{co,fit}-T_{co})/T_{co,fit}$.

Equation (1) fits $T_{co}(d_{Ag})$ for $d_{Pb}=3$ nm if we assume $\lambda_{Pb}=0.57$ as appropriate for quench condensed Pb films,[24] set $\beta$ to 0.30 and $T_o$ to 30.1 K. This value of $\beta$ lies between that derived from a free-electron model and that obtained from heat-capacity measurements.[25] Physically, $T_{co}(d_{Ag}=0)$, which is set by $T_o$, is the temperature at which strong pairing correlations on an isolated Pb grain appear. It should decrease with decreasing Pb grain radius due to finite-size effects[9,26,27] and thus it is less than the bulk Pb value. The high quality of the fit is expected, in so far as these bilayers fall in the Cooper limit. The quench condensation technique ensures the SN interfaces are clean. Also, $(d_{Pb}+d_{Ag})<20$ nm, which does not exceed either $\xi_S \simeq 60$ nm (Ref. 23) for quench condensed Pb, or $\xi_N$. We hasten to add that $T_{co}(d_{Ag})$ for $d_{Pb}=3$ nm does have a slight downward curvature, which can be fit by including a small repulsive interaction in the Ag, $\lambda_{Ag}=-0.017\pm0.004$ and setting $T_o=28.7$ K and $\beta=0.277$ [see Fig. 3].[1]

Surprisingly, $T_{co}(d_{Ag})$ of the thinner bilayers, $d_{Pb}=2.3$ and 1.5 nm, which should be deeper in the Cooper limit, systematically deviates from the exponential dependence at small $T_{co}$.[28] It is possible to roughly fit the data at the higher $T_{co}$ by adjusting $T_o$ and setting $\lambda_{Pb}$, $\lambda_{Ag}$, and $\beta$ to the values used for $d_{Pb}=3$ nm (solid lines in Fig. 3). The required changes in $T_o$ with $d_{Pb}$, $T_o(d_{Pb}=2.3\text{ nm})=27.6$ K, and $T_o(d_{Pb}=1.5\text{ nm})=23.5$ K can easily be ascribed to a reduction in $T_{co}$ of the individual Pb grains as they shrink.[26,27] The data at lower $T_{co}$, however, fall faster than exponentially. This characteristic is brought out in the inset of Fig. 3, where the fractional deviation of the data from the fits is shown. For the bilayer with the thinnest Pb layer, the deviation becomes greater than 50%.

Some possible explanations for the deviations can be ruled out. The deviations are significantly larger than the breadth of the $R(T)$ and thus variations in the definition of $T_{co}$ cannot account for them. They also appear at large enough $d_{Ag}$ that they cannot be attributed to changes occurring at the Pb-Ag interface that could influence $T_{co}$ (e.g., alloying). Finally, one might argue that Eq. (1) only holds for smooth continuous bilayers and deviations might be expected for more complicated geometries such as nanoscale arrays. The high quality of the fit to the $d_{Pb}=3$ nm data, however, counters this argument.

Alternatively, we suggest that as $T_{co}$ decreases and the order-parameter amplitude on the Pb grains decreases amplitude fluctuations grow and lead to deviations from mean-field behavior. These fluctuations reduce the average pairing interactions. In more detail, we estimate the order-parameter amplitude by the average number of Cooper pairs per grain, $N_{cp}\simeq\nu_{Pb}V\Delta$, where $V$ is the grain volume.[9] For $N_{cp}\gg1$ the probability, $P_{AS}$, that an electron entering a Pb grain has a pairing interaction with another electron within $\Delta$ of $E_F$ or, equivalently, Andreev scatters from the grain, is close to 1.[33] Mean-field treatments of the proximity effect implicitly presume $P_{AS}=1$ through the use of the factor $d_{Pb}/(d_{Pb}+\beta d_{Ag})$ to account for the reduction of the pairing interaction induced by the metal. As $T_{co}$ and $\Delta$ decrease and $N_{cp}$ approaches and falls below 1, fluctuations in $N_{cp}$ are expected and reduce $P_{AS}$ below 1. Consequently, the volume for pairing interactions falls faster with $d_{Ag}$ than the factor $d_{Pb}/(d_{Pb}+\beta d_{Ag})$ would imply, leading to a more rapid decrease in $T_{co}(d_{Ag})$.[32]

Simple estimates reveal that the nanoarrays with the lowest $T_{co}$'s fall in this fluctuation dominated regime. Using $V$ for the $d_{Pb}=3$ nm film and $\Delta=\Delta_{bulk}T_{co}/T_{co,bulk}$, we find that $N_{cp}\simeq1$ for the lowest $T_{co}$ film ($T_{co}=0.2$ K). This film lies close to the mean-field prediction. At comparable $T_{co}$, the $d_{Pb}=1.5$- and 2.3-nm films which have smaller grains and thus smaller $N_{cp}$ deviate from the Cooper limit theory. Presumably these films are firmly in the fluctuation dominated regime. For a rough measure of the fluctuations, the data point for the $d_{Pb}=1.5$ nm film at $d_{Ag}=7.5$ nm must be shifted to 9.0 nm or by about 20% for it to fall on the mean-field prediction. In other words, fluctuations prevent 20% of the Pb volume in this film from promoting pairing interactions.

It is interesting, in the light of recent predictions of a quantum SMT,[10,11] that the fluctuation effects tend to drive the arrays more rapidly toward a metallic state (i.e., $T_{co}=0$). Spivak, Zyuzin, and Hruska[11] predicted a critical $x_{sc}\sim|\lambda_N|\ln\{[1+(r_c-r)/r]/2\pi\}$ (Ref. 34) for the SMT where $r_c\simeq\xi_s$. By identifying $x_{sc}=d_{Pb}/(d_{Pb}+d_{Ag})$ and using $\lambda_{Ag}=-0.017$, as obtained from the fit in Fig. 3, we can estimate the critical value $(d_{Ag}/d_{Pb})_c$ for the SMT. For the $d_{Pb}=3$-nm film, $r=7.5$ nm$=(V)^{1/3}$, $(d_{Ag}/d_{Pb})_c=14$. Presuming that $r$ scales as $d_{Pb}$, $(d_{Ag}/d_{Pb})_c=10$ for the $d_{Pb}=1.5$-nm film. The predicted $(d_{Ag}/d_{Pb})_c$ is higher than the $(d_{Ag}/d_{Pb})$ at which strong downward curvature emerges in the data. This discrepancy may arise because the theory presumes that the superconducting grain spacing is much larger than $r$ and $r\simeq\xi_{bulk}$. In the experiments, $r\ll\xi_{bulk}$, which may lead to stronger fluctuation effects than expected.

In summary, quench condensation of ultrathin bilayer films of Pb and Ag has been used to fabricate arrays of SNS junctions with nanoscale dimensions. The array $T_{co}$'s decrease as the Pb grain concentration decreases in a manner quantitatively consistent with the Cooper limit theory of the proximity effect until the average number of Cooper pairs per Pb grain approaches 1. The ensuing, more rapid decrease in $T_{co}$ with decreasing Pb grain concentration is qualitatively consistent with the arrays approaching a superconductor-to-metal transition.

We have benefited from discussions with Boris Spivak, Marina Hruska, Tony Houghton, Brad Marston, Sean Ling, Jiufeng Tu, and Myron Strongin. We acknowledge the support of NSF grants NSF-DMR 980193 and NSF-DMR 0203608.